# Isospin Dependence in Fission Fragment Yields

S. Garg[1]*, B. Maheshwari[2], D. Choudhury[3] and A. K. Jain[1]
[1]Amity Institute of Nuclear Science & Technology, Amity University UP, Noida-201313, INDIA
[2]Department of Physics, Faculty of Science, University of Malaya, 50603 Kuala Lumpur, Malaysia
[3]Department of Physics, Indian Institute of Technology Ropar, Ropar-140001, INDIA
* email: swat90.garg@gmail.com

## Introduction

Nuclear fission is a very complex phenomenon. With the advancements in techniques like gammy ray spectroscopy and mass spectrometry, it is possible to measure the isotopic yields of fission fragments with the precision of one unit in $A$ and $Z$. At present, the neutron-induced fission is very important for energy applications and the fission fragment yields influence the total energy release, decay heat and production of long lived waste. The present data on isotopic and charge yields of fission fragments at energies ~2 MeV, relevant for future reactors, are very scarce and needed to be closely examined to have a better and complete understanding of fission. Therefore, in the present paper, we try to understand the mechanism behind the production of isotopic yields of fission fragments emitted in fast neutron induced fission of $^{238}$U using the idea of conservation of isospin.

The idea of using isospin conservation to obtain fission fragment distribution was first proposed by Jain et al. [1]. Theoretical claims of Lane and Soper [2] support the idea. This idea has been further expanded by us to calculate the relative yields of fission fragments emitted in heavy ion induced and thermal neutron induced fission [3-5]. Our results successfully reproduced the gross features of partition wise fission fragments distribution for reactions, $^{208}$Pb($^{18}$O, f), $^{238}$U($^{18}$O, f) and $^{245}$Cm($n_{th}$, f). In the present paper, we apply the concept of isospin conservation for the first time to fast neutron induced fission in $^{238}$U(n, f) from a recent measurement [6].

## Formalism

We use the same formalism in the present paper as used in Ref. [3-5]. First, we consider a fast neutron induced fusion-fission reaction,

$$n + {}^{238}U \rightarrow {}^{239}U \rightarrow F_1 + F_2 + q \quad (1)$$

where $F_1$ and $F_2$ correspond to the emitted fission fragments and $q$ is the total number of neutrons emitted in fission.

**Assignment of Isospin** – We start assigning isospin from left hand side of the reaction. The neutron has isospin $T = T_3 = 1/2$. In the ground state, the target $^{238}$U has isospin $T(^{238}U) = T_3(^{238}U)=27$. The isospin of the compound nucleus (CN) $^{239}$U adds up to give a unique value, $T(^{239}U) = T_3(^{239}U) = 27+1/2=27.5$.

We, further, introduce an auxiliary concept of residual compound nucleus (RCN) which is the CN remaining after the emission of $q$ number of neutrons. This RCN further breaks into two fragments $F_1$ and $F_2$,

$$CN \rightarrow RCN + q \rightarrow F_1 + F_2 \quad (2)$$

From the conservation of isospin, the isospin of RCN should satisfy the following conditions,

$$\left| T_{F_1} - T_{F_2} \right| \leq T_{RCN} \leq T_{F_1} + T_{F_2} \quad (3)$$

and, $\left| T_{CN} - q/2 \right| \leq T_{RCN} \leq T_{CN} + q/2 \quad (4)$

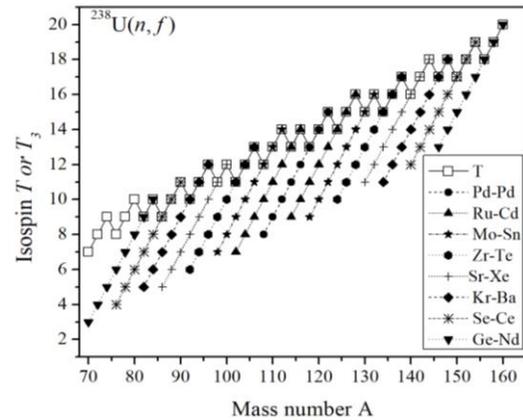

**Fig. 1** Isospin $T$ assigned to each mass number $A$ (open squares). The slanted lines gives the $T_3$ values of the fragments in various partitions.

The next important step is to assign isospin to various fission fragments. For this, we use Kelson's conjectures [7], as we have done in Ref. [3-5]. We consider three isobars corresponding to each mass number and assign the maximum isospin out of three $T_3$ values to that mass number. In this way, we make a complete and systematic assignment of isospin, see Fig. 1. We assign the isospin of $RCN$ as, $T_{RCN} = T_{F_1} + T_{F_2}$ with constraint implied by Eq. (4).

**Calculation of Relative Yields-** For each possible pair of fragments ($F_1$, $F_2$), emitted in $q$ $n$-emission channel in a particular partition, the isospin wave function of $RCN$ may be written as,

$$|T_{RCN}, T_{3RCN}\rangle_q = \langle T_{F_1} T_{F_2} T_{3F_1} T_{3F_2} | T_{RCN}, T_{3RCN} \rangle$$
$$|T_{F_1} T_{3F_1}\rangle |T_{F_2} T_{3F_2}\rangle \quad (5)$$

where $\langle T_{F_1} T_{F_2} T_{3F_1} T_{3F_2} | T_{RCN}, T_{3RCN} \rangle$ is the isospin C.G. coefficient (CGC). The yield of a particular fragment in a particular $n$-emission channel is given by the square of CGC's,

$$I_q = \langle T_{F_1} T_{F_2} T_{3F_1} T_{3F_2} | T_{RCN}, T_{3RCN} \rangle^2 = CGC^2 \quad (6)$$

The total yield of a particular fragment in a given partition can be written as,

$$I = \sum_q I_q = \sum_q CGC^2 \quad (7)$$

In a similar way, we calculate the yields of all the fragments in a given partition and then, we normalize the yields with respect to mass number having maximum yield to have the relative yields.

### Results and Discussion

The experimental data of $^{238}$U($n, f$) are available for the five partions namely, Mo-Sn, Zr-Te, Sr-Xe, Kr-Ba and Se-Ce [6]. However, there are no experimental data available for the weight factors corresponding to various $n$-emission channels. Therefore, we have considered the central three $n$-emission channels which correspond to the maximum yield of the fragments. In the present paper, we have shown the calculations for only two partitions, see Fig. 2. The calculations reproduce the relative isotopic yields of fission fragments reasonably well. However, the deviations may occur due to the presence of isomers or shell effects etc. as these are responsible for the fine structure features of fission fragment distribution. The overall agreement between the calculated and experimental data in various types of fission viz. heavy ion induced, thermal neutron induced and fast neutron-induced fission strongly supports the applicability and usefulness of conservation of isospin in neutron-rich nuclei. This gives us a completely new and different way to understand the isotopic yields of fission fragments in terms of isospin.

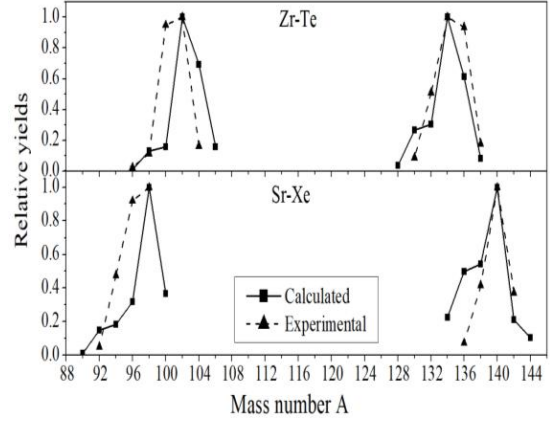

**Fig. 2** Relative yields of fission fragments emitted in $^{238}$U($n, f$) reaction vs. mass number $A$.

### Conclusion

This confirms the purity of isospin in neutron-rich systems even at high excitations in all types of fission and provides a unique tool for predicting the fission fragment mass distribution.